\documentstyle[12pt,fleqn,epsfig]{article}
\makeatletter
\renewcommand{\theequation}{\thesection.\arabic{equation}}
\@addtoreset{equation}{section}
\makeatother
\def\appendix#1{
  \addtocounter{section}{1}
  \setcounter{equation}{0}
  \renewcommand{\thesection}{\Alph{section}}
  \renewcommand{\theequation}{\Alph{section}.\arabic{equation}}
  \section*{Appendix \thesection\protect\indent #1}
  \addcontentsline{toc}{section}{Appendix \thesection #1}
  }
\newcommand{\beq}{\begin{equation}}
\newcommand{\eeq}{\end{equation}}
\newcommand{\beqy}{\begin{eqnarray}}
\newcommand{\eeqy}{\end{eqnarray}}

\def\tpi{\tilde{\pi}}

\def \ut#1{\rlap{\lower1ex\hbox{$\sim$}}#1{}}
\def \UT#1{\rlap{\lower1ex\hbox{\scriptsize$\sim$}}#1{}}

\newcommand{\LI}[1]{_{\mbox{ } #1}}

\def\ep{\epsilon}
\def\otep{\tilde{\epsilon}}

\def\CH{{\cal H}}
\def\CN{{\cal N}}
\def\CM{{\cal M}}
\def\CP{{\cal P}}
\def\M3{M^{(3)}}
\def\TM{\widetilde{M}}
\def\tx{\tilde{x}}

\begin{document}
\begin{flushright}
OU-HET/231\\gr-qc/9512017\\December 1995
\end{flushright}
\vspace{0.5in}
\begin{center}\Large{\bf A semiclassical interpretation
of the topological solutions for canonical quantum gravity}\\
\vspace{1cm}\renewcommand{\thefootnote}{\fnsymbol{footnote}}
\normalsize\ Kiyoshi Ezawa\footnote[1]
{Supported by JSPS.
e-mail address: ezawa@funpth.phys.sci.osaka-u.ac.jp}
        \setcounter{footnote}{0}
\vspace{0.5in}

        Department of Physics \\
        Osaka University, Toyonaka, Osaka 560, Japan\\
\vspace{0.1in}
\end{center}
\vspace{1.2in}
\baselineskip 17pt
\begin{abstract}

Ashtekar's formulation for canonical quantum gravity
is known to possess the topological solutions which
have their supports only on the moduli space $\CN$ of
flat $SL(2,{\bf C})$ connections.
We show that each point on the moduli space $\CN$ corresponds
to a geometric structure, or more precisely the
Lorentz group part of a family of Lorentzian
structures, on the flat (3+1)-dimensional spacetime.
A detailed analysis is given in the case where the spacetime
is homeomorphic to ${\bf R}\times T^{3}$.
Most of the points on the moduli space $\CN$ yield
pathological spacetimes which suffers from singularities on each
spatial hypersurface or which violates the
strong causality condition.
There is, however, a subspace of $\CN$ on which each point
corresponds to a family of regular spacetimes.

\end{abstract}

\vspace{0.2in}

\newpage

\baselineskip 20pt

\section{Introduction}

To quantize gravity is one of the most challenging problems
in physics. As one of the various approaches there is a program of
canonically quantizing general relativity.
The new variables discovered Ashtekar have simplified
the formulation of canonical quantum gravity \cite{ashte} and this
virtue of Ashtekar's new variables has served to find many new
aspects of nonperturbative canonical quantum gravity
\cite{schil}. One of these new aspects is that there exist
\lq topological solutions' for canonical quantum gravity
in terms of Ashtekar's new variables \cite{bren} \cite{kodama}.

In this paper we will focus on these topological solutions.
For this purpose it is convenient to use the chiral
Lagrangian formulation\cite{pleb}\cite{capo}\cite{baez}\cite{ueno}
which start with the following action:
\beq
-iI=\int_{M}(\Sigma^{i}\wedge F^{i}-\frac{\Lambda}{6}
\Sigma^{i}\wedge\Sigma^{i}), \label{eq:action}
\eeq
where $\Lambda$ is the cosmological constant and
$F^{i}=dA^{i}+\frac{1}{2}\ep^{ijk}A^{j}\wedge A^{k}$
is the curvature of the $SL(2,{\bf C})$ connection $A^{i}$,
which is related to the Levi-Civita spin-connection
$\{\omega^{\alpha\beta}\}$ as\footnote{The definition and properties
of the projector $P^{(-)i}_{\alpha\beta}$
are listed in Appendix A.}
\beq
A^{i}=-iP^{(-)i}\LI{\alpha\beta}\omega^{\alpha\beta}
=-(\frac{1}{2}\ep^{ijk}\omega^{jk}+i\omega^{0i}).
\label{eq:connection}
\eeq
$\Sigma^{i}$ is an $SL(2,{\bf C})$-valued two-form constructed from
the vierbein $\{e^{\alpha}\}$:
\beq
\Sigma^{i}=iP^{(-)i}\LI{\alpha\beta}e^{\alpha}\wedge e^{\beta}
=\frac{1}{2}\ep^{ijk}e^{j}\wedge e^{k}+ie^{0}\wedge e^{i}.
\label{eq:2form}
\eeq
This equation is equivalent to the following algebraic
constraint
\beq
\Sigma^{i}\wedge\Sigma^{j}=\frac{1}{3}\delta_{ij}
\Sigma^{k}\wedge\Sigma^{k},\label{eq:alcon}
\eeq
where the repeated indices are supposed to be contracted.
For simplicity we will restrict our analysis to the case where the
spacetime $M$ has the topology ${\bf R}\times\M3$ with $\M3$ being
a compact, oriented, 3 dimensional manifold without boundary.

We should note that the action (\ref{eq:action}) is nothing but
the BF action \cite{horo} with the gauge group $SL(2,{\bf C})$.
As a consequence the constraint equations which appear in
canonically quantized Ashtekar's formalism can be written in the
form of linear combinations of the constraint equations in the
BF theory. Therefore the solutions to the BF constraints are also
solutions to the Ashtekar constraints, which are called
the topological solutions.

There are two types of topological solutions according to whether
the cosmological constant $\Lambda$ vanishes or not. When $\Lambda\neq
0$ we have the unique solution, namely the
\lq\lq Chern-Simons solution"\cite{kodama}
\beqy
\Psi_{CS}[A^{i}_{a}]&=&\exp(iS_{CS}[A^{i}_{a}]),\nonumber \\
S_{CS}[A^{i}_{a}]&\equiv&i\frac{3}{2\Lambda}\int_{\M3}(A^{i}dA^{i}+
\frac{1}{3}\ep^{ijk}A^{i}\wedge A^{j}\wedge A^{k}).\label{eq:CS}
\eeqy
When $\Lambda$=0, on the other hand, we have the
topological solutions with their supports only on
flat connections \cite{bren}:
\beq
\Psi_{topo}[A^{i}_{a}]=\psi[A_{a}^{i}]\prod_{x\in\M3}\prod_{i,a}
\delta(\otep^{abc}F^{i}_{bc}(x)),\label{eq:topo}
\eeq
where we can consider the gauge
invariant functional $\psi[A^{i}_{a}]$ as
the function on the moduli space $\CN$ of flat connections modulo
identity-connected gauge transformations.

We are interested particularly in the geometrical significance of
these topological solutions, namely in what spacetimes correspond
to these solutions. It is known that we can give
a semiclassical interpretation to the Chern-Simons solution
\cite{kodama}\cite{soo}.
More precisely we can regard $S_{CS}[A^{i}_{a}]$ in eq.(\ref{eq:CS})
as a Hamilton principal functional which is subject to
the Hamilton-Jacobi equation, and we regard the WKB orbit,
namely the family of classical solutions extracted from the Hamilton
principal function $S_{CS}$, as corresponding to the quantum solution
$\Psi_{CS}[A^{i}_{a}]$. In the case of the Lorentzian signature
these WKB orbit yields locally (anti-)de Sitter spacetimes.

As for $\Psi_{topo}[A^{i}_{a}]$ in eq.(\ref{eq:topo}),
we do not know any works discussing on the relation to spacetimes,
except for the Euclidean case\cite{ueno2}.
We consider, however, that a semiclassical interpretation can be
given also to the topological solutions $\Psi_{topo}[A^{i}_{a}]$
in the sense explained below.
Formally the function $\psi[A^{i}_{a}]$ on the moduli space $\CN$
which appeared in eq.(\ref{eq:topo}) is considered to be a
superposition of the delta-function type wavefunctions
\beq
\psi_{n_{0}}(n)=\delta(n,n_{0}),
\eeq
where $n_{0}\in\CN$ and the argument $n$ ranges in $\CN$.
Finding the spacetimes
which correspond to the moduli $n_{0}\in\CN$ therefore amounts to
interpreting the topological solutions $\Psi_{topo}$ semiclassically.

This situation is similar to that of (2+1)-dimensional gravity
in the Chern-Simons form\cite{witt}. In (2+1)-dimensions the
moduli space of flat connections for
some non-compact gauge group is shown to correspond
to the space of geometric structures\cite{carl}, and
the spacetimes are explicitly constructed from the moduli
in the cases of relatively simple topologies with zero
cosmological constant \cite{carl2} \cite{louko} and with a nonzero
cosmological constant \cite{ezawa}. In particular in the case where
the cosmological constant vanishes, the moduli space of flat
spin connections yields the Lorentz group part of the Lorentzian
structure\footnote{For a detailed explanation of the geometric
structure, see e.g. \cite{gold}.}.

Because the BF theory is similar to the Chern-Simons gauge theory,
we expect that some of the techniques developed in
(2+1)-dimensional gravity can be applied also to (3+1)-dimensions
as far as the topological solutions are concerned.
We will see in this paper that this is indeed the case.
In particular, after imposing the reality conditions classically,
the moduli space of flat $SL(2,{\bf C})$ connections
turns out to yield the Lorentz group part of the (3+1)-dimensional
Lorentzian structures of the flat spacetime.
In the simplest case where the spatial
manifold $\M3$ has the topology
of a three-torus $T^{3}$ we will explicitly construct spacetimes from
the reduced phase space $\CM$ of the $SL(2,{\bf C})$ BF theory.
While most of the points on the moduli space $\CN$ correspond
to pathological spacetimes
which have singularities or which infringe
the strong causality condition, there is a subspace of
$\CN$ (with nonzero codimensions) each point on which yields a family
of well-behaved spacetimes.
Unlike in (2+1)-dimensions
the allowed values of the moduli of conjugate momenta (i.e.
$\Sigma^{i}$) are subject to severe restrictions under the condition
that the point in the reduced phase space should yield
spacetimes which are as well-behaved as possible.
Another definite difference is that some important information
which is necessary for constructing spacetimes is hidden
in the enlarged gauge transformation of the BF theory, i.e.
the Kalb-Ramond symmetry\cite{KR}. In (2+1)-dimensions
this is not the case because the corresponding enlarged gauge
transformation is almost equivalent to the diffeomorphism
transformation.

This paper is organized as follows.
In \S 2, after briefly reviewing the canonical formulation of the
BF theory with $\Lambda=0$, we construct the reduced phase space of
the $SL(2,{\bf C})$ BF theory on ${\bf R}\times T^{3}$.
Due to the non-compactness of the gauge group, there exist several
sectors each of which cannot be given solely by the total space
of the cotangent bundle over the moduli space.
In \S 3 we explicitly see the relation between the moduli space $\CN$
of flat $SL(2,{\bf C})$ connections and the (3+1)-dimensional
Lorentzian structures. In particular in the case with $\M3\approx
T^{3}$  we construct spacetimes with Lorentzian structures
whose projection to the Lorentz group is provided by
a point in the moduli space $\CN$.
\S 4 is devoted to the summary of the results and
the discussion on the extension to more general cases.
We briefly explain that the extension to the case $\M3\approx
T_{g}\times S^{1}$ is relatively tractable, where $T_{g}$ denotes
the Riemann surface of genus $g$.


\section{Reduced phase space of the $SL(2,{\bf C})$ BF theory}

In this section we look into the reduced phase space
of the $SL(2,{\bf C})$ BF theory. Our starting point is
the action (\ref{eq:action}) with vanishing cosmological constant
$\Lambda=0$:
$$
-iI=\int_{M}\Sigma^{i}\wedge F^{i}.
$$
Canonical formulation of this theory is obtained by
performing the (3+1)-decomposition of this action\footnote{We use
$t$ as the temporal coordinate and $(x^{a})=(x,y,z)$ as the
spatial coordinates.}. The result is
\beq
-iI=\int dt\int_{\M3}d^{3}x(\tpi^{ai}\dot{A}^{i}_{a}+A^{i}_{t}G^{i}
+\Sigma_{ta}^{i}\Phi^{ai}),\label{eq:bfcano}
\eeq
where we have set $\tpi^{ai}
\equiv\frac{1}{2}\otep^{abc}\Sigma^{i}_{bc}$ and $\dot{A}=
\frac{\partial}{\partial t}A$. Note that the temporal components
$A^{i}_{t}$ and $\Sigma^{i}_{ta}$ of the fields have no kinetic
terms. As a result these temporal components are regarded as
Lagrange multipliers with respect to which the variation of the
action yields first class constraints.
This system involves two types of
first class constraints. Gauss' law constraint
\beq
G^{i}\equiv D_{a}\tpi^{ai}\label{eq:gauss}
\eeq
generates under the Poisson brackets
$SL(2,{\bf C})$ gauge transformations
\beqy
\delta_{\theta}\Sigma&=&[\theta,\Sigma]\nonumber \\
\delta_{\theta}A&=&-D\theta\equiv-d\theta-[A,\theta],
\label{eq:gauge}
\eeqy
where we have set $\Sigma\equiv\Sigma^{i}J_{i}$ and $A\equiv
A^{i}J_{i}$.\footnote{
$J_{i}$ ($i=1,2,3$) denote $SL(2,{\bf C})$ generators which are
subject to the commutation relations:
$[J_{i},J_{j}]=\ep^{ijk}J_{k}$.}
$\theta=\theta^{i}J_{i}$ is an $SL(2,{\bf C})$-valued scalar field
on $\M3$. The remaining constraint
\beq
\Phi^{ai}\equiv \frac{1}{2}\otep^{abc}F^{i}_{bc}\label{eq:kr}
\eeq
generates the(generalized) Kalb-Ramond transformations\cite{KR}
\beqy
\delta_{\phi}\Sigma&=&-D\phi\equiv
-d\phi-A\wedge\phi-\phi\wedge A\nonumber \\
\delta_{\phi}A&=&0,\label{eq:KR}
\eeqy
where $\phi=\phi^{i}J_{i}$ is an $SL(2,{\bf C})$-valued one form.
These Kalb-Ramond transformations are known to include the
diffeomorphisms as a special case\cite{horo}\cite{ueno}.

The reduced phase space (or the physical phase space)
is defined as the quotient space of the constraint surface
modulo gauge transformations in a broad sense.
The constraint surface is the subspace of the original phase space
$\{(A^{i}_{a},\tpi^{ai})\}$ which satisfies the first class
constraint equations $G^{i}=\Phi^{ai}=0$. The gauge
transformations in a broad sense are the transformations
generated by the first class constraints. In our case they
consist of the small $SL(2,{\bf C})$ gauge transformations and the
Kalb-Ramond transformations.
By the argument in ref.\cite{horo},
the reduced phase space is considered
in general to be the total space
of the cotangent bundle $T^{\ast}\CN$ over the moduli space of flat
$SL(2,{\bf C})$ connections modulo small gauge transformations,
namely the gauge transformations which is homotopic to the identity
\footnote{As is seen in the next subsection, this is not necessarily
the case if the moduli space $\CN$ includes null rotations.}.
The base space $\CN$ and the fiber $T^{\ast}_{n_{0}}\CN$ are
respectively coordinatized by the gauge equivalent classes
of connections $A^{i}_{a}$ and 2-forms $\Sigma^{i}_{ab}$
which are subject to the constraint equations.

As is mentioned in ref.\cite{horo}, the moduli space $\CN$ in general
consists of disconnected sectors which are related with each other
by large gauge transformations, namely the gauge transformations
which cannot be continuously connected to the identity.
This fact has an interesting effect in the quantum BF theory
such as e.g. the appearance of $\theta$-states which
also appear in the Yang-Mills
theory. However, we are now interested in the semiclassical
interpretation of the moduli space. The $SL(2,{\bf C})$ connections
which are related with each other by an $SL(2,{\bf C})$
gauge transformation are expected to yield the same spacetime
metric. Thus we will henceforth concentrate only on the sector
$\CN_{0}$ which is connected to the trivial connection $A^{i}_{a}=0$.

In order to paremetrize $\CN_{0}$, it is convenient to use holonomies
along the non-contractible loops:
\beq
H(\alpha)\equiv h_{\alpha}[0,1]\equiv
\CP\exp(\int_{0}^{1}ds\dot{\alpha}^{a}(s)A_{a}),
\eeq
where $\alpha:[0,1]\rightarrow\M3$ ($\alpha(0)=\alpha(1)=x_{0}$)
is a loop and $\CP$ stands for the
path ordering with smaller $s$ to the
left. Because the connection $A_{a}$ is flat the holonomy depends
only on the homotopy class $[\alpha]$ of the loop $\alpha$:
\beq
H(\alpha)=H(\alpha^{\prime})\equiv H[\alpha]\quad\mbox{if}\quad
[\alpha]=[\alpha^{\prime}].
\eeq
The (small) gauge transformation $A_{a}(x)\rightarrow g(x)A_{a}(x)
g^{-1}(x)+g(x)\partial_{a}g^{-1}(x)$ on the connection $A_{a}(x)$
is translated into the conjugation by $g(x_{0})$ on the holonomy
\beq
H[\alpha]\rightarrow g(x_{0})H[\alpha]g^{-1}(x_{0}).
\eeq
The moduli space $\CN_{0}$ is therefore identical to the space
of equivalence classes of homomorphisms from the fundamental group
$\pi_{1}(\M3)$ to the group $SL(2,{\bf C})$ modulo conjugations:
\beq
\CN_{0}={\rm Hom}(\pi_{1}(\M3),SL(2,{\bf C}))/\sim.
\eeq
In the following we will explicitly construct $\CN_{0}$ and the
reduced phase space by exploiting the holonomy in the case where
the spatial manifold $\M3$ is homeomorphic to the three torus
$T^{3}$.

\subsection{The reduced phase space on $T^{3}$}

Let us now consider the case with $\M3\approx T^{3}$.
We first construct the moduli space $\CN_{0}$.
It follows from $\pi_{1}(T^{3})\cong{\bf Z}\oplus{\bf Z}\oplus
{\bf Z}$ that the holonomies evaluated along the generators
$\{\alpha,\beta,\gamma\}$ of $\pi_{1}(T^{3})$ commute mutually.
By taking appropriate conjugations we find that the moduli
space $\CN_{0}$ consists of several \lq disconnected' sectors:
\beq
\CN_{0}=\CN_{S}\oplus\sigma_{n_{1},n_{2},n_{3}\in\{0,1\}}\oplus
\CN_{F}^{n_{1},n_{2},n_{3}}.
\eeq
The standard sector $\CN_{S}$ is characterized by
\beqy
H[\alpha]&=&\exp((u+ia)J_{1})\nonumber \\
H[\beta]&=&\exp((v+ib)J_{1})\nonumber \\
H[\gamma]&=&\exp((w+ic)J_{1}),\label{eq:nst}
\eeqy
where $u,v,w$ are real numbers defined modulo $4\pi$ and
$a,b,c\in{\bf R}$. The gauge equivalence under
$g(x_{0})=\exp(\pi J_{2})$ further imposes the following
equivalence condition
\beq
(a,b,c;u,v,w)\sim -(a,b,c;u,v,w).\label{eq:Z2}
\eeq
As a consequence the standard sector $\CN_{S}$ has the topology
$(T^{3}\times{\bf R}^{3})/{\bf Z}_{2}$.

The flat sectors $\CN_{F}^{n_{1},n_{2},n_{3}}$ are parametrized
by the following holonomies:
\beqy
H[\alpha]&=&(-1)^{n_{1}}\exp(\xi(J_{2}+iJ_{1}))\nonumber \\
H[\beta]&=&(-1)^{n_{2}}\exp(\eta(J_{2}+iJ_{1}))\nonumber \\
H[\gamma]&=&(-1)^{n_{3}}\exp(\zeta(J_{2}+iJ_{1})),\label{eq:nf}
\eeqy
where $\xi,\eta,\zeta$ are complex numbers which do not vanish
simultaneously. The gauge equivalence under
$g(x_{0})=\exp(-i\kappa J_{3})$ with $\kappa\in{\bf C}$ tells
us that $(\xi,\eta,\zeta)$ provide the homogeneous coordinates
on ${\bf CP}^{2}$:
\beq
(\xi,\eta,\zeta)\sim e^{\kappa}(\xi,\eta,\zeta).\label{eq:CP2}
\eeq
Thus we find $\CN_{F}^{n_{1},n_{2},n_{3}}\approx{\bf CP}^{2}$.
Bcause two flat sectors $\CN_{F}^{n_{1},n_{2},n_{3}}$ and
$\CN_{F}^{n_{1}^{\prime},n_{2}^{\prime},n_{3}^{\prime}}$
with $(\delta n_{1},\delta n_{2},\delta n_{3})=
(n_{1}-n_{1}^{\prime},n_{2}-n_{2}^{\prime},
n_{3}-n_{3}^{\prime})\neq(0,0,0)$ are related with each other by
the large gauge transformation $g(x)=\exp(2\pi(\delta n_{1}x+
\delta n_{2}y+\delta n_{3}z)J_{3})$, they are considered as
corresponding
to the same set of the spacetimes. Thus in the following
we will consider only one flat sector $\CN_{F}\equiv\CN_{F}^{0,0,0}$.

Let us next construct the reduced phase space, more precisely,
the sectors of the reduced phase space which are associated
with the sectors $\CN_{S}$ and $\CN_{F}$ of the moduli space
$\CN$ of flat $SL(2,{\bf C})$ connections.

For this purpose we have to find the $SL(2,{\bf C})$ connection
which gives the holonomies (\ref{eq:nst}) or (\ref{eq:nf}).
This task is relatively easy in the present case. By
making an appropriate gauge choice and an adequate choice of
periodic coordinates $(x,y,z)$ with period $1$,
we find the desired connections.
A connection which belongs to $\CN_{S}$ is given by
\beq
A_{a}dx^{a}=[(u+ia)dx+(v+ib)dy+(w+ic)dz]J_{1}\equiv
(dW_{1}+idW_{2})J_{1}=dWJ_{1},\label{eq:NST}
\eeq
and a connection lying in $\CN_{F}$ is
\beq
A_{a}dx^{a}=(\xi dx+\eta dy+\zeta dz)(J_{2}+iJ_{1})
\equiv d\Xi(J_{2}+iJ_{1})=(d\Xi_{1}+id\Xi_{2})(J_{2}+iJ_{1}).
\label{eq:NF}
\eeq

Once we have obtained the explicit forms of the connection,
we can construct the whole reduced phase space
by solving Gauss' law constraint (\ref{eq:gauss}) for
the conjugate momenta $\tpi^{ai}$, or equivalently for
the two form $\Sigma^{i}_{ab}$, and by gauging away
the extra degrees of freedom by means of the Kalb-Ramond symmetry
(\ref{eq:KR}). The calculation is somewhat tedious and is
illustrated in Appendix B. Here we describe the result only.

The two-form associated with the connection (\ref{eq:NST})
is given by:
\beq
\Sigma=(\lambda_{1}dy\wedge dz+\lambda_{2}dz\wedge
dx+\lambda_{3}dx\wedge dy)J_{1},\label{eq:TNST}
\eeq
where $\lambda_{1},\lambda_{2},\lambda_{3}$ are complex numbers.
By the same reasoning which lead us to eq.(\ref{eq:Z2}),
the rigid gauge transformation $g(x)=\exp(\pi J_{2})$
tells us the equivalence relation
\beq
(A,\Sigma)\sim-(A,\Sigma)
\eeq
with $A$ and $\Sigma$ given by (\ref{eq:NST}) and (\ref{eq:TNST})
respectively. Thus we expect the sector of the reduced
phase space in question to have a cotangent bundle structure.
We can see that this is indeed the case by investigating
the reduced action which is obtained by substituting
eqs.(\ref{eq:NST}) and (\ref{eq:TNST}) into the BF action
(\ref{eq:bfcano}):
\beq
-iI^{\ast}_{\CN_{S}}=\int dt[\lambda_{1}(\dot{u}+i\dot{a})
+\lambda_{2}(\dot{v}+i\dot{b})+\lambda_{3}(\dot{w}+i\dot{c})].
\label{eq:NSac}
\eeq
This is exactly the symplectic potential
which takes the form of the holomorphic cotangent vector
defined on $\CN_{S}$. For a point in $\CN_{S}$
the fiber coordinate $(\lambda_{1},\lambda_{2},\lambda_{3})$
ranges over
the whole space of ${\bf C}^{3}$. Thus the sector $\CM_{S}$ of the
reduced phase space associated with $\CN_{S}$ turns out to be
the (holomorphic) cotangent bundle $T^{\ast}_{\CH}((T^{3}\times
{\bf R}^{3})/{\bf Z}_{2})$.\footnote{Precisely speaking
the cotangent bundle structure breaks down at the points
$A=2\pi(n_{1}dx+n_{2}dy+n_{3}dz)J_{1}$ in $\CN_{S}$, where
$n_{1},n_{2},n_{3}$ take their values in $\{0,1\}$. For a detail
see Appendix B.}

The generic expression of the two-forms associated with $\CN_{F}$
is
\beqy
\Sigma&=&(pdy\wedge dz+qdz\wedge dx+rdx\wedge dy)
\frac{J_{2}-iJ_{3}}{2} \nonumber \\
& &\qquad
+(Pdy\wedge dz+Qdz\wedge dx+Rdx\wedge dy)\frac{J_{2}+iJ_{3}}{2},
\label{eq:TNF}
\eeqy
where $p,q,r$ and $P,Q,R$ are all complex numbers and $p,q,r$ are
subject to the following constraint
\beq
p\xi+q\eta+r\zeta=0.\label{eq:CNF}
\eeq
The same rigid gauge transformation $g(x)=\exp(-i\kappa J_{3})$
as that which leads to the equivalence relation (\ref{eq:CP2})
yields in turn the following equivalence relation:
\beq
(p,q,r;P,Q,R)\sim(e^{-\kappa}p,e^{-\kappa}q,e^{-\kappa}r;
e^{\kappa}P,e^{\kappa}Q,e^{\kappa}R),
\eeq
which holds at the same time with eq.(\ref{eq:CP2}).
As is easily seen
this sector $\CM_{F}$ of the reduced phase space is not
equal to the cotangent bundle over $\CN_{F}$ because
the complex dimensions of $\CN_{F}$ is 2 while that of the space
of two-forms (\ref{eq:TNF}) associated with a point in $\CN_{F}$
is 5. We can nevertheless see that the symplectic potential
of the original phase space is inherited also to the
sector $\CM_{F}$. By substituting
eqs.(\ref{eq:NF}) and (\ref{eq:TNF}) into eq.(\ref{eq:bfcano})
and by taking account of the constraint (\ref{eq:CNF}),
we find the reduced action
\beq
-iI^{\ast}_{\CN_{F}}=\int[p\dot{\xi}+q\dot{\eta}+r\dot{\zeta}
-\dot{\kappa}(t)(p\xi+q\eta+r\zeta)].\label{eq:NFac}
\eeq
The first three terms in the r.h.s yields the symplectic potential.
Owing to the constraint (\ref{eq:CNF}) this becomes the well-defined
holomorphic cotangent vector on $\CN_{F}\approx{\bf CP}_{2}$.
Thus the space which is coordinatized by $(\xi,\eta,\zeta;p,q,r)$
is exactly the  holomorphic cotangent bundle $T^{\ast}_{\CH}
{\bf CP}^{2}$ over the moduli space $\CN_{F}$. There remains in fact
extra three-dimensional space $\{(P,Q,R)\}$ which is isomorphic
to ${\bf C}^{3}$. The whole sector $\CM_{F}$
of the reduced phase space
associated with $\CN_{F}$ are considered to be the vector bundle
over ${\bf CP}^{2}$ with the fiber being the direct product
$((T^{\ast}_{\CH})_{n_{0}}{\bf CP}^{2})
\times{\bf C}^{3}$. Because structure
functions for the fiber ${\bf C}^{3}$ are provided by the ratios
$\xi^{\prime}/\xi$ of different holomorphic coordinates,
this \lq\lq vector bundle" is equivalent to the direct product
$(T^{\ast}_{\CH}{\bf CP}^{2})\times{\bf C}^{3}$.

Before ending this section we briefly
investigate the classical dynamics
on the reduced phase space. The action (\ref{eq:NSac}) implies
that the dynamics on $\CM_{S}$ is trivial. Thus the classical
solutions are given by the connection and two-forms
which are gauge equivalent to the connection (\ref{eq:NST}) and
the two-form (\ref{eq:TNST}) with time-independent parameters.
On the other hand the classical solutions for the
action (\ref{eq:NFac}) appear to be somewhat nontrivial:
\beqy
(\xi,\eta,\zeta)(t)&=&e^{\kappa(t)}(\xi_{0},\eta_{0},\zeta_{0})
\nonumber \\
(p,q,r)(t)&=&e^{-\kappa(t)}(p_{0},q_{0},r_{0}).
\eeqy
A scrutiny shows that the above time evolution is nothing but the
time-dependent gauge transformation with $g(t,x)=\exp(-i\kappa(t)
J_{3})$. We can thus consider the classical solutions to be gauge
equivalent to the connection (\ref{eq:NF}) and the two-form
(\ref{eq:TNF}) with time-independent parameters
$(\xi,\eta,\zeta:p,q,r)$. We cannot determine the
evolution of the parameters $(P,Q,R)$ from the reduced action
(\ref{eq:NFac}) and so these parameters may appear to be
arbitrary functions of time $t$. However, if we use equations of
motion $(D\Sigma)_{tab}=0$, $(P,Q,R)$ turns out to be gauge
equivalent to a constant vector in ${\bf C}^{3}$.


\section{Relation to (3+1)-dimensional Lorentzian structures}

\subsection{General framework}

Now that we have constructed the moduli space $\CN_{0}$,
let us consider what spacetimes correspond to each point on $\CN_{0}$.
By analogy with (2+1)-dimensions \cite{carl2}\cite{ezawa},
we expect that the Kalb-Ramond transformation (\ref{eq:KR})
with a time-dependent parameter $\phi$ plays an important role
when we construct non-degenerate spacetime metric from
the solutions to the constraints which are
obtained in the previous section.

There are however two essential differences compared to the
(2+1)-dimensional case. One is the presence of the algebraic
constraint (\ref{eq:alcon}), owing to which the allowed values
of the parameters in eq.(\ref{eq:TNST}) or eq.(\ref{eq:TNF})
are considered to be restricted.
The other comes from the extra symmetries in the BF theory compared
to those in Ashtekar's formalism. In (2+1)-dimensions the
translation part of the $ISO(2,1)$ gauge symmetry corresponds to
the Kalb-Ramond symmetry in (3+1)-dimensional BF theory.
This translation part is shown to be equivalent to diffeomorphisms
under on-shell and when the dreibein is non-degenerate \cite{witt},
whereas in (3+1)-dimensions the Kalb-Ramond transformation
contains extra symmetry other than diffeomorphisms\cite{ueno2}.
Thus in (3+1)-dimensions it is possible that a
Kalb-Ramond transformation relates two spacetimes which are
not diffeomorphic with each other. This implies
that some important information on the spacetime is
involved in the Kalb-Ramond transformation (\ref{eq:KR}).
In (2+1)-dimensions this was not the case; the translation part of
the $ISO(2,1)$ gauge transformation at most changes the appearance
of the singular part of the spacetime\cite{louko}.

By analogy to the (2+1)-dimensional case it would be standard
to take the following recipe for spacetime construction: i) prepare
the solution $(A,\Sigma)$ obtained in the previous section; ii)
take an appropriate time-dependent Kalb-Ramond transformation
(\ref{eq:KR}); iii) impose the algebraic constraint
(\ref{eq:alcon}) and classical reality conditions which are
equivalent to declaring that
\beq
A^{i}=-iP^{(-)i}_{\quad\alpha\beta}\omega^{\alpha\beta},
\quad\Sigma^{i}=iP^{(-)i}_{\alpha\beta}e^{\alpha}\wedge
e^{\beta}\label{eq:real}
\eeq
with a real spin-connection $\omega^{\alpha\beta}$ and a real
vierbein $e^{\alpha}$; and iv) construct the metric
from the vierbein $e^{\alpha}$ and give a spacetime geometrical
interpretation to the corresponding point $(A,\Sigma)$ in the
reduced phase space.

To carry out this procedure is in principle possible but in
practice difficult. We will therefore
take a converse procedure, namely: i) prepare the connection
$A\in\CN_{0}$; ii) extract the spin connection
$\omega^{\alpha\beta}$ by imposing the classical reality condition
(\ref{eq:real}) on $A$; iii) find the vierbein $e^{\alpha}$ w.r.t.
which $\omega^{\alpha\beta}$ is torsion-free
\beq
de^{\alpha}+\omega^{\alpha}\LI{\beta}\wedge e^{\beta}=0;
\label{eq:torsionfree}
\eeq
iv) investigate the properties of the spacetime obtained from the
vierbein; v) construct the two-form $\Sigma^{i}$ by using
the second equation of eq.(\ref{eq:real}); and vi) transform
$\Sigma^{i}J_{i}$ into a standard form by using the Kalb-Ramond
symmetry (\ref{eq:KR}).

Let us now explore the concrete relation between the moduli
space $\CN_{0}$ and the space of Lorentzian structures.
We first consider the generic case $M\approx{\bf R}\times\M3$.
By combining the flatness of the $SL(2,{\bf C})$ connection
and the reality condition (\ref{eq:real}), we see that
the spin connection $\omega^{\alpha\beta}$ is flat.
Thus, on the universal covering $\TM\approx{\bf R}\times
\widetilde{\M3}$ which is defined in a similar manner as
that in Appendix B, we can express $\omega^{\alpha\beta}$
in the form of a pure gauge:
\beq
\omega^{\alpha}\LI{\beta}(\tx^{\mu})=
(\Lambda^{-1}(\tx^{\mu}))^{\alpha}\LI{\gamma}
d\Lambda(\tx^{\mu})^{\gamma}\LI{\beta},
\eeq
where $\Lambda(\tx^{\mu})^{\alpha}\LI{\beta}\equiv
(\CP\exp\int_{0}^{\tx^{\mu}}\omega(\tilde{y}^{\nu}))^{\alpha}
\LI{\beta}$ is the integrated spin connection\footnote{
We mean by $x^{\mu}$ and $\tx^{\mu}$ the point on $M$ and
that on $\TM$ respectively.}. Using this
equation, the torsion-free condition (\ref{eq:torsionfree})
is cast into the closedness condition
$$
d[\Lambda(\tx^{\mu})^{\alpha}
\LI{\beta}e^{\beta}(\tx^{\mu})]=0.
$$
Because $\TM$ is simply-connected
by definition, the torsion-free condition is completely solved by
\beq
e^{\alpha}(\tx^{\mu})=(\Lambda^{-1}(\tx^{\mu}))^{\alpha}\LI{\beta}
dX^{\beta}(\tx^{\mu}),\label{eq:master1}
\eeq
where the set $\{X^{\alpha}(\tx^{\mu})\}$ is considered to be the
embedding of $\TM$ into (the universal covering of) an adequate
subspace of the (3+1)-dimensional Minkowski space $M^{3+1}$.
In order for the vierbein $e^{\alpha}(\tx^{\mu})$ to be well-defined
on $M$, it must satisfy the \lq periodicity condition'
\beq
e^{\alpha}(\gamma+\tx^{\mu})=e^{\alpha}(\tx^{\mu})\quad\mbox{for}
\quad^{\forall}[\gamma]\in\pi_{1}(M),\label{eq:peri1}
\eeq
(plus some conditions necessary when $\TM$ is not contractible
to a point).
By substituting eq.(\ref{eq:master1}) into eq.(\ref{eq:peri1})
we find
$$
dX^{\alpha}(\gamma+\tx^{\mu})=d(H_{0}[\gamma]^{\alpha}\LI{\beta}
X^{\beta}(\tx^{\mu})),
$$
where $H_{0}[\gamma]^{\alpha}\LI{\beta}$ stands for the holonomy
of the spin connection $\omega^{\alpha}\LI{\beta}$ evaluated
along the loop $\gamma$.
By integrating this we obtain the important result:
\beq
X^{\alpha}(\gamma+\tx^{\mu})=H_{0}[\gamma]^{\alpha}\LI{\beta}
X^{\beta}(\tx^{\mu})+V[\gamma]^{\alpha},\label{eq:Lorentzian}
\eeq
which states that the periodicity condition of the embedding
functions $\{X^{\alpha}(x^{\mu})\}$ is given by Poincar\'{e}
transformations which are isometries of the Minkowski space.
The consistency of the periodicity conditions associated with
all the loops in $\pi_{1}(M)\cong\pi_{1}(\M3)$, it is necessary that
the set of Poincar\'{e} transformations
$\{(H_{0}[\gamma]^{\alpha}\LI{\beta},V[\gamma]^{\alpha})|\gamma\in
\pi_{1}(\M3)\}$
should give a homomorphism from $\pi_{1}(\M3)$ into the
(3+1)-dimensional Poincar\'{e} group. This is precisely a
Lorentzian structure on the manifold $M\approx{\bf R}\times\M3$
\cite{gold}.
Thus we can say that the moduli of the flat $SL(2,{\bf C})$
connections on a spacetime manifold $M$ specifies the
Lorentz transformation part of the geometric structures
each of which belongs to
$$
{\rm Hom}(\pi_{1}(\M3),(\CP^{3+1})^{\uparrow}_{+})/\sim,
$$
where $(\CP^{3+1})^{\uparrow}_{+}$ denotes the
proper orthochronous Poincar\'{e}
group in (3+1)-dimensions and $\sim$ stands for the equivalence under
the conjugation by (proper orthochronous)
Poincar\'{e} transformations.

(Remark) In order for a Lorentzian structure to provide a physically
permissible spacetime, it must satisfy several conditions.
First it must act on an appropriate subspace of $M^{3+1}$ properly
discontinuously in order to avoid the degeneracy of the spacetime.
This can usually be resolved by considering the universal
covering of the subspace.
Second, in order not to render spatial manifolds $\M3$ to collapse,
it must be embedded into the subgroup of
the Poincar\'{e} group which is of rank 3.\footnote{
The rank here is meant to be the maximal number of mutually commuting
generators in the subgroup.}
The third condition, the spacelike nature of its action
on a relevant region of $M^{3+1}$,
is necessary for the spatial hypersurface $\M3$ to be spacelike.
We will see, however, this third condition is not sufficient
for the strong causality condition to hold.


\subsection{Application to $M\approx{\bf R}\times T^{3}$}

In this subsection we will apply the
procedure explained in the previous
subsection to the case where $\M3$ is homeomorphic to $T^{3}$.
We should bear in mind that, because
$\pi_{1}(T^{3})\cong{\bf Z}\oplus{\bf Z}\oplus{\bf Z}$,
the Lorentz structure under consideration is generated by
three Poincar\'{e} transformations which mutually commute.

First we consider the case with $A=dWJ_{1}\in\CN_{S}$.
After imposing the classical reality condition
the integrated spin connection is given by
\beq
(\Lambda^{\alpha}\LI{\beta})=\left(\begin{array}{cccc}
\cosh W_{2}&-\sinh W_{2}& 0 & 0 \\
-\sinh W_{2}&\cosh W_{2}& 0 & 0 \\
0 & 0 & \cos W_{1} & -\sin W_{1} \\
0 & 0 & \sin W_{1} & \cos W_{1}
\end{array}\right),
\eeq
with $(W_{1},W_{2})\equiv(ux+vy+wz,ax+by+cz)$. The four cases need
to be investigated separately.\footnote{
Because $SO(3,1)^{\uparrow}$ is obtained from $SL(2,{\bf C})$ by
neglecting the overall sign factor $\pm1$, we will consider in this
section that $u,v,w$ are defined modulo $2\pi$.}

I) For $(u,v,w)\neq\vec{0}\neq(a,b,c)$, the equation
(\ref{eq:master1}) is rewritten as
\beqy
\cosh W_{2}e^{0}-\sinh W_{2}e^{1}&=&dX^{0}\nonumber \\
-\sinh W_{2}e^{0}+\cosh W_{2}e^{1}&=&dX^{1}\label{eq:01} \\
\cos W_{1}e^{2}-\sin W_{1}e^{3}&=&dX^{2}\nonumber \\
\sin W_{1}e^{2}+\cos W_{1}e^{3}&=&dX^{3}.\label{eq:23}
\eeqy
We choose an adequate set $(X^{\alpha})$ by imposing the
condition that the vierbein $e^{\alpha}$ is single-valued on
$M$ and yields the metric with a correct signature. A candidate
which yields the most admissible spacetime is
\beq
(X^{\alpha})=(\tau\cosh(W_{2}+\alpha),-\tau\sinh(W_{2}+\alpha),
Z\sin(W_{1}+\beta),-Z\cos(W_{1}+\beta)),\label{eq:NN}
\eeq
where the time function $\tau$ is single-valued on $M$ and
$\alpha$ is an arbitrary single-valued function on $M$.
$\beta$ is such that $(\cos\beta,\sin\beta)$ is single-valued
on $M$ and $Z$ is a single-valued function on $M$
which is bounded away from zero.
The spacetime metric constructed from the solution to the
eqs.(\ref{eq:01})(\ref{eq:23}) with $(X^{\alpha})$ given by
(\ref{eq:NN}) takes the following form
\beq
ds^{2}=\eta_{\alpha\beta}e^{\alpha}e^{\beta}=
-d\tau^{2}+\tau^{2}d(W_{2}+\alpha)^{2}+dZ^{2}+
Z^{2}d(W_{1}+2\pi\vec{n}\cdot\vec{x}+\beta_{0})^{2},\label{eq:nn}
\eeq
where we have made a decomposition: $\beta=2\pi\vec{n}\cdot\vec{x}
+\beta_{0}$ with $\beta_{0}$ a single-valued function on $M$ and
$\vec{n}\cdot\vec{x}\equiv nx+my+lz$ ($n,m,l\in{\bf Z}$).
We can easily see that the embedding (\ref{eq:NN}) gives the
spacetime whose Lorentzian structure is given by the corresponding
holonomy (\ref{eq:nst}) without any translation part.

We should note that this Lorentzian structure is singular
because it is embedded into the rank 2 subgroup of
the Poincar\'{e} group. We therefore expect that
the spacetime also possesses singularities. We will see that
this is indeed the case. For simplicity
we set $(W_{1},W_{2})=(\frac{\pi}{2}x,y)$, $\alpha=\beta=0$
and $Z=\frac{1}{\pi}(3+\cos2\pi z)$.
In this case the metric (\ref{eq:nn}) becomes
\beq
ds^{2}=-d\tau^{2}+\tau^{2}dy^{2}+4\sin^{2}2\pi zdz^{2}
+\frac{1}{4}(3+\cos z)^{2}dx^{2}.\label{eq:snn}
\eeq
Obviously the singularities exist at $z=0$ and $z=\frac{1}{2}$
in which the metric degenerates.
By looking into the vierbein we find that these two singularities
divide the spacetime into two regions whose local Lorentz
frames have the opposite orientations. While such spacetime
is not permitted physically, we can represent this spacetime
by using two Minkowski spaces. We will
depict in Fig.1 how the $d\tau=dW_{2}=0$ section of $M$
is realized by the two $(X^{2},X^{3})$-planes in $M^{3+1}$.
We see that the resulting $d\tau=dW_{2}=0$ section takes the
form of the \lq\lq double covering of the Klein-bottle" which
is homeomorphic to $T^{2}$. The spacetimes with arbitrary
functions $(Z,\alpha,\beta)$ and with generical parameters
$(u,v,w;a,b,c)$ also suffer from the same type of singularities
and so they
cannot be considered to be physical, at least classically.

\begin{figure}[t]
\begin{center}
\epsfig{file=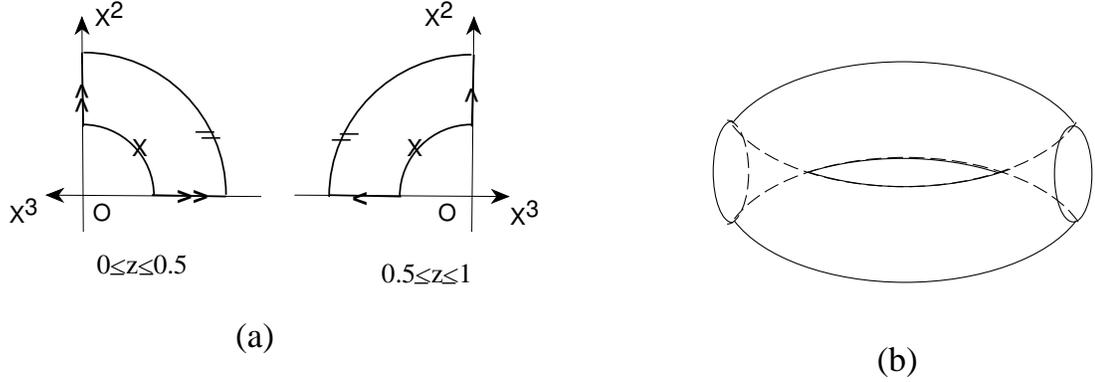,height=7cm}
\end{center}
\caption{The $d\tau=dW_{2}=0$ section of the spacetime (3.12).
By identifying the segments (or the arcs)
marked by the same symbol in (a), we obtain the
\lq\lq double-covering of the Klein-bottle" as shown in (b).}
\end{figure}

We can now construct the two-form $\Sigma=\Sigma^{i}J_{i}$
from the vierbein which is obtained by solving eqs.(\ref{eq:01})
(\ref{eq:23}). The result is
\beqy
\Sigma&=&D\phi \\
\phi&=&\frac{1}{2}
[-i\tau^{2}d(W_{2}+\alpha)+Z^{2}d(W_{1}+\beta)]J_{1}\nonumber \\
&+&
[i\tau\cosh\alpha d(Z\sin\beta)-\tau\sinh\alpha d(Z\cos\beta)
\nonumber \\
& &+\tau Z(i\cosh\alpha\cos\beta+\sinh\alpha\sin\beta)dW_{1}]J_{2}
\nonumber \\
&+&
[-i\tau\cosh\alpha d(Z\cos\beta)-\tau\sinh\alpha d(Z\sin\beta)
\nonumber \\
& &+\tau Z(i\cosh\alpha\sin\beta-\sinh\alpha\cos\beta)dW_{1}]J_{3},
\nonumber
\eeqy
namely the two-form in this case takes the form of a pure Kalb-Ramond
transformation. Thus we see that some of the information
(such as $Z$ and $\beta$) which is necessary
to construct the spacetime is hidden in the
Kalb-Ramond symmetry and that the only allowed values of the
fiber-coordinates are $\lambda_{1}=\lambda_{2}=\lambda_{3}=0$.

II) For $(a,b,c)=\vec{0}\neq(u,v,w)$, we can embed the Lorentzian
structure into a rank 3 subgroup of the Poincar\'{e} group
$\{R_{23},T^{0},T^{1}\}$, where $R_{ij}$ and $T^{\alpha}$
respectively denote the rotation in the $(X^{i},X^{j})$-plane
and the translation in the $X^{\alpha}$-direction. This subgroup,
however, includes time-translations. The corresponding spacetimes
thus involve timelike-tori and so
they are not considered to be physical.
Of course we could use the Lorentzian structure which is embedded in
the rank 2 subgroup $\{R_{23},T^{1}\}$. But in this case the
singularities similar to that appeared in case I) always exist in
the resulting spacetime. This case therefore corresponds to
a set of spacetimes not allowed in general relativity.

III) For $(u,v,w)=\vec{0}\neq(a,b,c)$, the situation drastically
changes. The Lorentzian structure in this case can be embedded
into the rank 3 subgroup $\{L_{1},T^{2},T^{3}\}$, where $L_{i}$
stands for the Lorentz boost in the $(X^{0},X^{i})$-plane.
Fortunately the action of this subgroup on the region $\{(X^{0})^{2}-
(X^{1})^{2}>0\}\in M^{3+1}$ is spacelike. This indicates that
this case corresponds to a set of well-behaved spacetimes.

More concretely, since $W_{1}=0$ in this case, eq.(\ref{eq:23})
is merely the closedness condition on $(e^{2},e^{3})$. A choice
of the embedding which yield well-behaved spacetimes is:
$$
(X^{\alpha})=(\tau\cosh(W_{2}+\alpha),-\tau\sinh(W_{2}+\alpha),
\vec{\beta}\cdot\vec{x}+\psi_{2},\vec{\gamma}\cdot\vec{x}+\psi_{3}),
$$
where $\vec{\beta}$ and $\vec{\gamma}$ are constant vectors in
${\bf R}^{3}$ and
$\psi_{2}$ and $\psi_{3}$ are single-valued functions
on $M$. Substituting this into eqs.(\ref{eq:01})(\ref{eq:23}),
we find
\beqy
e^{0}&=&d\tau\cosh\alpha+\tau\sinh\alpha d(W_{2}+\alpha)\nonumber \\
e^{1}&=&-d\tau\sinh\alpha-\tau\cosh\alpha
d(W_{2}+\alpha) \nonumber \\
e^{2}&=&dX^{2}=\vec{\beta}\cdot d\vec{x}+d\psi_{2} \nonumber \\
e^{3}&=&dX^{3}=\vec{\gamma}\cdot d\vec{x}+d\psi_{3}.
\eeqy
This vierbein gives a physically permissible spacetime whose only
pathology is the initial singularity at $\tau=0$:
\beq
ds^{2}=-d\tau^{2}+\tau^{2}d(W_{2}+\alpha)^{2}+(dX^{2})^{2}+
(dX^{3})^{2}.
\eeq
In this case the two-form cannot be gauged away completely
\beqy
\Sigma&=&dX^{2}\wedge dX^{3}J_{1}+D\phi  \\
\phi&=&-\frac{i}{2}\tau^{2}d(W_{2}+\alpha)J_{1}+
(\tau\sinh\alpha dX^{3}+i\tau\cosh\alpha dX^{2})J_{2}
\nonumber \\ & &
-(\tau\sinh\alpha dX^{2}-i\tau\cosh\alpha dX^{3})J_{3}. \nonumber
\eeqy
Now the parameter space
$(\lambda_{1},\lambda_{2},\lambda_{3})\in{\bf C}^{3}$ is
restricted to ${\bf R}^{3}\backslash\{(0,0,0)\}$
by the condition that it corresponds to
physically accepted spacetimes. While these parameters
$(\lambda_{i})$ contain some information on $(X^{2},X^{3})$,
we cannot extract all the information
on $(X^{2},X^{3})$ from them. The
remaining information is still hidden in the Kalb-Ramond
transformation.

IV) For $(a,b,c)=(u,v,w)=\vec{0}$. In this case also we can embed
the geometric structure into the rank 3 subgroup
$\{T^{1},T^{2},T^{3}\}$, whose action on the whole Minkowski space
is spacelike. We expect this case to correspond to the spacetimes
without any singularity. Indeed the well-behaved solutions for
eqs.(\ref{eq:01})(\ref{eq:23}) are given by
\beqy
e^{0}&=&dX^{0}=dT \nonumber \\
e^{1}&=&dX^{1}=\vec{\alpha}\cdot d\vec{x}+d\psi_{1} \nonumber \\
e^{2}&=&dX^{2}=\vec{\beta}\cdot d\vec{x}+d\psi_{2} \nonumber \\
e^{3}&=&dX^{3}=\vec{\gamma}\cdot d\vec{x}+d\psi_{3},
\eeqy
where $\vec{\alpha},\vec{\beta},\vec{\gamma}\in{\bf R}^{3}$
are constant vectors and $\psi_{i}$ ($i=1,2,3$) are single-valued
functions on $M$.
The metric constructed from this vierbein is non-singular
\beq
ds^{2}=-dT^{2}+dX^{i}dX^{i}.
\eeq
As in case III) the two-form in this case cannot be gauged away
completely
\beq
\Sigma=\frac{1}{2}\ep^{ijk}dX^{i}\wedge dX^{j}J_{k}+d(iTdX^{i}J_{i}).
\eeq
Besides, we can extract all the important information on the
spacetime from the reduced phase space. This is considered to be
characteristic to the  origin $0\in\CN_{0}$ in the case $\M3\approx
T^{3}$.

Next we construct spacetimes corresponding to $A=d\Xi(J_{2}+iJ_{1})
\in\CN_{F}$. After imposing the classical reality condition,
we obtain the integrated spin-connection
\beq
(\Lambda^{\alpha}\LI{\beta})=\left(\begin{array}{cccc}
1+\frac{1}{2}\Xi\overline{\Xi}& -\Xi_{1} &
-\Xi_{2} & -\frac{1}{2}\Xi\overline{\Xi} \\
-\Xi_{1} & 1 & 0 & \Xi_{1} \\
-\Xi_{2} & 0 & 1 & \Xi_{2} \\
\frac{1}{2}\Xi\overline{\Xi} & - \Xi_{1} &
-\Xi_{2} & 1-\frac{1}{2}\Xi\overline{\Xi}
\end{array}\right),
\eeq
where the bar denotes the complex conjugation.
Substituting this expression into eq.(\ref{eq:master1}),
we find the following equations
\beqy
e^{0}-e^{3}&=&d(X^{0}-X^{3}) \nonumber \\
e^{1}-\Xi_{1}(e^{0}-e^{3})&=&dX^{1} \nonumber \\
e^{2}-\Xi_{2}(e^{0}-e^{3})&=&dX^{2} \nonumber \\
e^{0}+e^{3}-2\Xi_{1}e^{1}-2\Xi_{2}e^{2}
+\Xi\overline{\Xi}(e^{0}-e^{3})&=&d(X^{0}+X^{3}).
\label{eq:nullmaster}
\eeqy
We have to solve these equations taking the periodicity
condition (\ref{eq:peri1}) into account.
We consider the two cases separately.

I)' For ${\rm Re}(\vec{\xi})\neq\vec{0}\neq{\rm Im}(\vec{\xi})$,
with $\vec{\xi}\equiv(\xi,\eta,\gamma)$, the Lorentzian structure
in question is embedded into the subgroup $\{N^{1+},N^{2+},T^{+}\}$,
where $T^{+}$ is the translation in the $X^{+}(\equiv X^{0}+X^{3})$-
direction and $N^{\hat{i}+}$ ($\hat{i}=1,2$) is the null-rotation
which stabilizes $X^{+}$ and $X^{\hat{i}}$. Because this subgroup
contains translation in the null-direction, there is a possibility
that the closed null curves appear.

A candidate
for the embedding function $(X^{\alpha})$ which yields the most
well-behaved spacetime is given by
\beqy
(X^{-},X^{1},X^{2},X^{+})&=&(e^{t},-e^{t}\Xi_{1}+\Phi_{1},
-e^{t}\Xi_{2}+\Phi_{2},\nonumber \\
& & -e^{-t}+Z_{+}+\Phi_{3}+e^{t}\Xi\overline{\Xi}
-2\Xi_{1}\Phi_{1}-2\Xi_{2}\Phi_{2}),
\eeqy
where $\Phi_{i}$ are arbitrary single-valued functions on $M$,
$t$ is a time function and
$Z_{+}\equiv\vec{\alpha}_{+}\cdot\vec{x}$
with $\vec{\alpha}_{+}$ being a constant vector in ${\bf R}^{3}$.
By substituting this into eq.(\ref{eq:nullmaster}), we find
\beqy
(e^{-},e^{1},e^{2},e^{+})
&=&(e^{t}dt,-e^{t}d\Xi_{1}+d\Phi_{1},-e^{t}d\Xi_{2}+d\Phi_{2},
\nonumber \\
& & e^{-t}dt+d(Z_{+}+\Phi_{3})-2\Phi_{1}d\Xi_{1}
-2\Phi_{2}d\Xi_{2}),
\eeqy
where we have set $e^{\pm}\equiv e^{0}\pm e^{3}$.
In order to investigate the corresponding spacetime it is
sufficient to consider the case $\Phi_{i}=0$ ($i=1,2,3$).
As is shown in Appendix C, while the resulting metric
\beq
ds^{2}=-dt^{2}-e^{t}dtdZ_{+}+e^{2t}d\Xi d\overline{\Xi}
\label{eq:nullmetric}
\eeq
may or may not have closed causal curves\footnote{
In fact this metric has at worst closed null curves.},
it certainly violates the strong causality condition\cite{hawk}.
The corresponding spacetime is therefore not so desirable in
general relativity.

As for the two-form $\Sigma$ constructed from the above vierbein,
we can show that it
takes the form of a pure Kalb-Ramond transformation
$D\phi$ with $\phi$ given by a complicated $SL(2,{\bf C})$-valued
one-form
involving $t$ and $dZ_{+}$. Thus only the origin of the fiber
$T^{\ast}_{n_{0}}\CN_{F}\times {\bf C}^{3}$ (cf. eq.(\ref{eq:TNF}))
is relevant.

II)' For ${\rm Im}(\vec{\xi})=0\neq{\rm Re}(\vec{\xi})$
(or equivalently for ${\rm Re}(\vec{\xi})=\vec{0}\neq{\rm Im}
(\vec{\xi})$), the Lorentzian structure is embedded into
the rank 3 subgroup $\{N^{2+},T^{+},T^{2}\}$. Also in this case
the spacetime is expected to infringe the strong causality
condition. This is indeed the case. By substituting the embedding
function $(X^{-},X^{1},X^{2},X^{+})=(e^{t},-e^{t}\Xi_{1},
\vec{\beta}\cdot\vec{x},-e^{-t}+Z_{+}+e^{t}(\Xi_{1})^{2})$
into eq.(\ref{eq:nullmaster}) with $\Xi_{2}=0$ and by
using the resulting solution, we find the following metric
\beq
ds^{2}=-dt^{2}-e^{t}dtdZ_{+}+e^{2t}(d\Xi_{1})^{2}+
(\vec{\beta}\cdot d\vec{x})^{2},
\eeq
which violates the strong causality condition in the same way
that eq.(\ref{eq:nullmetric}) does. The two-form $\Sigma$ in this
case takes the following form
\beq
\Sigma=-id(\vec{\beta}\cdot\vec{x})\wedge dZ_{+}J_{+}+D\phi,
\eeq
with $\phi$ being an $SL(2,{\bf C})$-valued one-form involving
$d(\vec{\beta}\cdot\vec{x})$, $dZ_{+}$ and $t$. Some important
information on the spacetime is hidden in the Kalb-Ramond
transformation also in this case.
The allowed region in the fiber $T^{\ast}_{n_{0}}\CN_{F}
\times{\bf C}^{3}$ is restricted to $\{0\}\times[(i{\bf R})^{3}
\backslash\{(0,0,0)\}]$. The cotangent space therefore
does not play an essential role in constructing
spacetimes from a point in $\CM_{F}$.


\section{Summary and discussion}

In this paper we have made an attempt to give a semiclassical
interpretation to the topological solutions for canonical quantum
gravity by elucidating the relationship between the moduli
space $\CN_{0}$ of flat $SL(2,{\bf C})$ connections and
the space of Lorentzian structures on the flat (3+1)-dimensional
spacetime with a fixed topology ${\bf R}\times\M3$.
We have shown that, after imposing the classical reality condition,
a point on the moduli space $\CN_{0}$ gives a unique
Lorentz transformation part of the Lorentzian structures.
Thus the topological solution of the form $\delta(n,n_{0})$
corresponds to a family of spacetimes each of which has a
Lorentzian structure whose projection onto the Lorentz group
is specified by the holonomy group which determines
the point $n_{0}\in\CN_{0}$.
In the case of $\M3\approx T^{3}$, we have explicitly
constructed the spacetimes corresponding to each point
on the moduli space $\CN_{0}=\CN_{S}\oplus\CN_{F}$.
While most of the points correspond to physically
undesirable spacetimes which have singularities or which
violate the strong causality condition, a subspace
of $\CN_{S}$ yields spacetimes which are physically
well-behaved. Each sectors $\CM_{S}$ and $\CM_{F}$
of the reduced phase space were
regarded as the total space of the fiber bundle
over the moduli space $\CN_{S}$ or $\CN_{F}$ which contains
the holomorphic cotangent bundle as a subspace.
The allowed region of each fiber seems to be restricted
by requiring that the points in it correspond
to spacetimes which are as physical as possible.
It appears that this restriction is relaxed when the rank
of the holonomy group is smaller than its maximal
value on each sector.
In any case we have seen some of the important information
on the spacetime metric are hidden in the Kalb-Ramond
transformation, with the only exception being the case
where the holonomy of the spin connection is trivial.

In order to establish that these results hold also to
the cases of more generical topologies, more profound
acquaintance with topology and geometric structures
of 3-dimensional manifolds is required and thus this problem is
left to the future investigation.

We can, however, assert that there exist a subspace of
the moduli space $\CN_{0}$ of flat $SL(2,{\bf C})$ connections
which corresponds to a family of
physically well-behaved spacetimes, at least when $\M3$ has the
topology $T_{g}\times S^{1}$ with $T_{g}$ being a 2-dimensional
Riemann surface of genus $g\geq2$. We know that,
in (2+1)-dimensions, there exist Lorentzian structures on the
spacetime ${\bf R}\times T_{g}$ which is well-defined on
the domain of dependence \cite{witt} \cite{mess}
\beq
(X^{0})^{2}-(X^{1})^{2}-(X^{2})^{2}>0.
\eeq
Now $\pi_{1}(T_{g}\times S^{1})$ is isomorphic
to $\pi_{1}(T_{g})\oplus{\bf Z}$
and the (2+1)-dimensional Poincar\'{e} group can be
naturally embedded into the (3+1)-dimensional Poincar\'{e} group.
These facts tell us that there is at least a set of
physically well-defined Lorentzian structures on ${\bf R}\times T_{g}
\times S^{1}$, each of which consists of a (2+1)-dimensional
Lorentzian structure on ${\bf R}\times T_{g}$ which acts
on $X^{3}=const.$ hypersurfaces and a translation
in the $X^{3}$-direction which yields the periodicity condition
for the $S^{1}$-direction. Because the moduli space $\CN_{0}$
gives only the Lorentz transformation part, the moduli
of the flat spin connections which corresponds
to the above Lorentzian structure does not
have any information on the structure of the $S^{1}$-direction.
It would be interesting to investigate whether
these Lorentzian structures can be extended into
more complicated structures or not.

The relation between Ashtekar's formalism and the $SL(2,{\bf C})$
BF theory can be extended to $N=1,2$ supergravities \cite{ezawa2}.
The existence of the topological solutions for these
super-extended versions of Ashtekar's formalism \cite{mats}
\cite{sano} \cite{shira} is understood as a natural consequence of
this relation.
It is of interest to explore what spacetimes correspond to
these topological solutions for supergravities because
in supergravities the non-vanishing torsion in general
gives some influence on the spacetime geometry.

\vspace{0.2in}

\noindent Acknowledgments

I would like to thank Prof. K. Kikkawa,
Prof. H. Itoyama and H. Kunitomo
for useful discussions and careful readings of the manuscript.


\setcounter{section}{0}
\setcounter{section}{0}


\appendix{The projector $P^{(-)i}_{\alpha\beta}$}

Here we provide the definition and the properties of the
projector $P^{(-)i}_{\alpha\beta}$. First we define the
projection operator $P^{(-)\alpha\beta}_{\quad\gamma\delta}$
into the space of anti-self-dual Lorentz bi-vectors:
\beq
P^{(-)\alpha\beta}_{\quad\gamma\delta}=
\frac{1}{4}(\delta^{\alpha}_{\gamma}\delta^{\beta}_{\delta}-
\delta^{\alpha}_{\delta}\delta^{\beta}_{\gamma}
-i\ep^{\alpha\beta}_{\quad\gamma\delta}),
\eeq
where $\ep^{\alpha\beta\gamma\delta}$ is the totally anti-symmetric
pseudo tensor with $\ep^{0123}=\ep^{123}=1$. We use the
metric $(\eta_{\alpha\beta})=(\eta^{\alpha\beta})=
{\rm diag}(-1,1,1,1)$ in raising or lowering the Lorentz indices.
This projection operator possesses the following properties
\beq
P^{(-)\alpha\beta}_{\quad\gamma\delta}=
-\frac{i}{2}\ep^{\alpha\beta}_{\quad\alpha^{\prime}\beta^{\prime}}
P^{(-)\alpha^{\prime}\beta^{\prime}}_{\quad\gamma\delta}=-
\frac{i}{2}P^{(-)\alpha\beta}_{\quad\gamma^{\prime}\delta^{\prime}}
\ep^{\gamma^{\prime}\delta^{\prime}}_{\quad\gamma\delta}=
P^{(-)\alpha\beta}_{\quad\alpha^{\prime}\beta^{\prime}}
P^{(-)\alpha^{\prime}\beta^{\prime}}_{\quad\gamma\delta}.
\eeq
The projector $P^{(-)i}_{\quad\alpha\beta}$ is defined as
\beqy
P^{(-)i}_{\quad\alpha\beta}&\equiv&
\frac{1}{2}(\delta^{0}_{\alpha}\delta^{i}_{\beta}-
\delta^{0}_{\beta}\delta^{i}_{\alpha}-i
\ep^{0i}_{\quad\alpha\beta})\nonumber \\
&=&2P^{(-)0i}_{\quad\alpha\beta}=-i\ep^{ijk}
P^{(-)jk}_{\quad\alpha\beta}.
\eeqy
This projector satisfies the following identities
\beqy
P^{(-)i}_{\quad\gamma\delta}P^{(-)i\alpha\beta}&=&
-P^{(-)\alpha\beta}_{\quad\gamma\delta} \\
\eta^{\beta\delta}P^{(-)i}_{\quad\alpha\beta}
P^{(-)j}_{\quad\delta\gamma}
&=&\frac{i}{2}\ep^{ijk}P^{(-)k}_{\quad\alpha\gamma}+
\frac{1}{4}\delta^{ij}\eta_{\alpha\gamma}.
\eeqy
Using this projector we can give the relation between $SO(3,1)$
representation $\Lambda^{\alpha}\LI{\beta}$ and $SO(3,{\bf C})$
representation $\Lambda^{ij}$ of the (proper orthochronous)
Lorentz group:
\beq
\Lambda^{ij}=-P^{(-)i}_{\quad\alpha\beta}\Lambda^{\alpha}
\LI{\gamma}\Lambda^{\beta}\LI{\delta}P^{(-)j\gamma\delta}.
\eeq
This $SO(3,{\bf C})$ representation is obtained as the adjoint
representation of $SL(2,{\bf C})$:
\beq
(e^{\theta^{k}J_{k}})^{ij}\Phi^{j}J_{i}=e^{\theta^{k}J_{k}}
\Phi^{j}J_{j}e^{-\theta^{k}J_{k}},
\eeq
where $(J_{k})^{ij}=\ep^{ikj}$ are the $SL(2,{\bf C})$ generators
in the adjoint representation.


\appendix{Solutions to the constraint $D_{a}\tpi^{ai}=0$}

Let us start by providing the formal solution
in the generic case. We first notice that the constraint equation
$G^{i}=D_{a}\tpi^{ai}=0$ is equivalent to the restriction of
the following equation to a spatial hypersurface $\M3${}
\beq
D\Sigma=d\Sigma+A\wedge\Sigma-\Sigma\wedge A=0.\label{eq:kr2}
\eeq

Next we introduce the universal covering $\widetilde{\M3}$ which is
the space of all the homotopy classes of the curves
in $\M3$ which starts from, say, the origin $x=0$. We will
decompose the point $\tx$ on $\widetilde{\M3}$ as
$\tx=\gamma+x$, where
$\gamma+x$ denotes the curve which first passes
along the loop $\gamma$ beginning at the origin $x=0$
and then goes from
the origin to the point $x$ on $\M3$ by way of the shortest
path measured by some positive-definite background metric
on $\M3$. If there are more than one shortest paths we will
choose one by some continuous scheme. We will also denote by
$\gamma+\tx$ the homotopy class of the path which first
goes along the loop $\gamma$ and then goes from the origin to
the point $x\in\M3$ along a path representing $\tx$. We should
note that several relations hold such as
\beqy
\gamma\cdot\gamma^{\prime}+\tx&=&
\gamma+(\gamma^{\prime}+\tx)\nonumber \\
\gamma^{-1}+(\gamma+\tx)&=&\tx,\nonumber
\eeqy
but that in general $\gamma^{\prime}+(\gamma+\tx)\neq\gamma+
(\gamma^{\prime}+\tx)$.

Let us now solve eq.(\ref{eq:kr2}). Because
any flat connection on $\widetilde{\M3}$ is written as a pure gauge
\beq
A^{ij}\equiv\ep^{ikj}A^{k}=(\Lambda^{-1})^{ik}d\Lambda^{kj},
\eeq
eq.(\ref{eq:kr2}) on $\widetilde{\M3}$ is equivalent to
the following equation
\beq
d(\Lambda^{ij}\Sigma^{j})=0.
\eeq
Assuming that $H_{2}(\widetilde{\M3})$ is trivial
the above equation is completely solved by
\beq
\Sigma^{i}=(\Lambda^{-1})^{ij}d\Phi^{j}=
D[(\Lambda^{-1})^{ij}\Phi^{j}],\label{eq:master2}
\eeq
where $\Phi^{i}(\tx)$ is a one-form on $\widetilde{\M3}$.
In order for $\Sigma^{i}$ to be well-defined (single-valued) on
$\M3$, we further have to impose the \lq periodic condition'
\beq
\Sigma^{i}(\gamma+\tx)=\Sigma^{i}(\tx)\quad \mbox{for}\quad
^{\forall}[\gamma]\in\pi_{1}(\M3,0),\label{eq:peri2}
\eeq
and some other conditions necessary in the case where
$\widetilde{\M3}$
is non-contractible to a point. Using eq.(\ref{eq:master2}),
the condition (\ref{eq:peri2}) can be rewritten in terms of
$\Phi^{i}$:
\beq
\Phi^{i}(\gamma+\tx)=H_{0}[\gamma]^{ij}\Phi^{j}(\tx)
+\Delta\Phi^{i}(\gamma;\tx),\label{eq:b1}
\eeq
where $H_{0}[\gamma]$ is the holonomy of $A^{ij}$ along
the loop $\gamma$ and $\{\delta\Phi^{i}_{\gamma}(\tx)\}$
is a set of closed
one-forms on $\widetilde{\M3}$ which is subject to the relation
\beq
\Delta\Phi^{i}(\gamma\cdot\gamma^{\prime};\tx)=
H_{0}[\gamma]^{ij}\Delta\Phi^{j}(\gamma^{\prime};\tx)
+\Delta\Phi^{i}(\gamma;\gamma^{\prime}+\tx).\label{eq:b2}
\eeq

Here we make a remark. It appears from eq.(\ref{eq:master2})
that all the $\Sigma^{i}$ can be gauged away by using the
Kalb-Ramond transformation (\ref{eq:KR}). But this is not
necessarily the case. In order for the solution (\ref{eq:master2})
to be gauged away, it is necessary for the one-form
$(\Lambda^{-1})^{ij}\Phi^{j}$ to be well-defined on $\M3$.
This cannot follow only from the conditions (\ref{eq:b1})
(\ref{eq:b2}). Thus in general the solution (\ref{eq:master2})
cannot be gauged away completely. We will explicitly see
this in the case with $\M3\approx T^{3}$.

\subsection{Solutions on the three-torus}

We first investigate the case where the connection
belongs to the standard sector $\CN_{S}$, namely $A=dWJ_{1}$
as is seen in eq.(\ref{eq:NST}). The integrated connection
$\Lambda^{ij}$ is given by
\beq
(\Lambda^{ij})=\left(\begin{array}{ccc}
1 & 0 & 0 \\ 0 & \cos W & -\sin W \\ 0 & \sin W & \cos W
\end{array}\right).
\eeq
Eq.(\ref{eq:master2}) is thus rewritten as
\beqy
\Sigma^{1}&=&d\Phi^{1}\label{eq:B1} \\
\Sigma^{2}&=&\cos Wd\Phi^{2}+\sin Wd\Phi^{3}\label{eq:B2} \\
\Sigma^{3}&=&-\sin Wd\Phi^{2}+\cos Wd\Phi^{3}.\label{eq:B3}
\eeqy
Let us now solve these equations taking account of the condition
(\ref{eq:peri2}).  Eq.(\ref{eq:B1}) is easily solved by
\beq
\Sigma^{1}=\lambda_{1}dy\wedge dz+\lambda_{2}dz\wedge dx
+\lambda_{3}dx\wedge dy+d\phi^{1},\label{eq:S1}
\eeq
where $\lambda_{1},\lambda_{2},\lambda_{3}$ are constant complex
numbers and $\phi^{1}$ is an arbitrary one-form on $\M3$.
In order for $(\Sigma^{2},\Sigma^{3})$ to be single-valued on
$\M3$, we must choose the form of $(\Phi^{2},\Phi^{3})$ as:
\beq
\left\{\begin{array}{lll}
\Phi^{2}&=&B\cos(W+\alpha)dx+C\cos(W+\beta)dy\\
\quad& &\quad+D\cos(W+\gamma)dz
+E\cos(W+\delta)d\Phi\\
\Phi^{3}&=&B\sin(W+\alpha)dx+C\sin(W+\beta)dy \\
\quad& &\quad+D\sin(W+\gamma)dz
+E\sin(W+\delta)d\Phi,
\end{array}\right.
\eeq
where $B,C,D,E,\Phi$ are arbitrary single-valued function
on $\M3$ and $\alpha,\beta,\gamma,\delta$ are scalar
functions on $\widetilde{\M3}$ such that
$(\cos\alpha,\sin\alpha)$ and the similar expressions with
$\alpha$ replaced by $\beta,\gamma,\delta$ are single-valued
on $\M3$. By substituting these expressions into eqs.(\ref{eq:B2})
(\ref{eq:B3}), we find
\beq
\left\{\begin{array}{lll}
\Sigma^{2}&=&d\Phi^{\prime 2}-dW\wedge\Phi^{\prime 3}\\
\Sigma^{3}&=&d\Phi^{\prime 3}+dW\wedge\Phi^{\prime 2},\label{eq:S2}
\end{array}\right.
\eeq
where $(\Phi^{\prime 2},\Phi^{\prime 3})\equiv(B\cos\alpha dx+
C\cos\beta dy+D\cos\gamma dz+E\cos\delta d\Phi,B\sin\alpha dx+
C\sin\beta dy+D\sin\gamma dz+E\sin\delta d\Phi)$ are well-defined
one-forms on $\M3$. By putting the equations (\ref{eq:S1})
(\ref{eq:S2}) into together, we obtain the final result:
\beq
\Sigma=\Sigma^{i}J_{i}=(\lambda_{1}dy\wedge dz+\lambda_{2}dz\wedge
dx+\lambda_{3}dx\wedge dy)J_{1}
+D[\phi^{1}J_{1}+\Phi^{\prime 2}J_{2}+\Phi^{\prime 3}J_{3}].
\eeq
By using the Kalb-Ramond transformation, we see that this is
gauge-equivalent to eq.(\ref{eq:TNST}).

A special consideration is needed for the case with
$W=2\pi(n_{1}x+n_{2}y+n_{3}z)$ ($n_{1},n_{2},n_{3}\in\{0,1\}$).
In this case we can set, for example, $W+\alpha=0$ by choosing
$\alpha=-W$. Thus the choice $d\Phi^{\hat{i}}=
\lambda^{\hat{i}}_{1}dy\wedge dz+\lambda^{\hat{i}}_{2}dz\wedge dx+
\lambda^{\hat{i}}_{3}dx\wedge dy$ ($i=2,3$) also yields single-valued
$\Sigma^{\hat{i}}$ ($\hat{i}=2,3$). The resulting two-forms are gauge
equivalent to the conjugation classes of
general $SL(2,{\bf C})$-valued de Rham cohomology classes, namely,
\beqy
e^{2\pi(n_{1}x+n_{2}y+n_{3}z)J_{1}}\Sigma
e^{-2\pi(n_{1}x+n_{2}y+n_{3}z)J_{1}}=(\lambda_{1}^{i}dy\wedge
dz+\lambda_{2}^{i}dz\wedge dx+
\lambda_{3}^{i}dx\wedge dy)J_{i},\nonumber
\eeqy
where $(\lambda_{a}^{i})$ belongs to the space
$GL(3,{\bf C})/SO(3,{\bf C})$ with complex dimension 6.
\footnote{We should note that $GL(3,{\bf C})$ here denotes the
Lie algebra while $SO(3,{\bf C})$ stands for the Lie group.}

Next we will solve eq.(\ref{eq:kr2}) in the case where
$A=d\Xi(J_{2}+iJ_{1})\in\CN_{F}$.
In this case eq.(\ref{eq:master2}) becomes
\beq
\Sigma^{i}J_{i}=e^{-2WJ_{+}}(d\Phi^{+}J_{+}+d\Phi^{-}J_{-}+
d\Phi^{3}J_{3})e^{2WJ_{+}},
\eeq
where $J_{\pm}\equiv\frac{1}{2}(J_{2}\pm iJ_{1})$ are \lq null
basis' of the generators of $SL(2,{\bf C})$. In components
we find
\beqy
\Sigma^{-}&=&d\Phi^{-}\label{eq:B4} \\
\Sigma^{3}&=&d\Phi^{3}-iWd\Phi^{-}\label{eq:B5} \\
\Sigma^{+}&=&d\Phi^{+}+W^{2}d\Phi^{-}+2iWd\Phi^{3}.\label{eq:B6}
\eeqy
Eq.(\ref{eq:B4}) tells us that $\Sigma^{-}$ is an ordinary
closed two-form on $\M3$:
\beq
\Sigma^{-}=\Sigma^{-}_{0}+d\phi^{-},\label{eq:S4}
\eeq
where $\Sigma^{-}_{0}\equiv pdy\wedge dz+qdz\wedge dx+rdx\wedge dy$
and $\phi^{-}$ is an arbitrary one-form on
$\M3$. By substituting this into eq.(\ref{eq:B5}), taking
its exterior derivative, and by comparing the terms appeared in
the obtained equation, we find
\beqy
dW\wedge\Sigma^{-}_{0}&=&0\label{eq:S5}\\
\Sigma^{3}&=&idW\wedge\phi^{-}+\Sigma^{3}_{0}+d\phi^{3},
\label{eq:S6}
\eeqy
where $\Sigma^{3}_{0}$ is a linear combination of $\{dy\wedge dz,
dz\wedge dx,dx\wedge dy\}$ and $\phi^{3}$ is an arbitrary one-form
on $\M3$. In order to determine the form of $\Sigma^{+}$ we
substitute all the obtained results into eq.(\ref{eq:B6}) and
take the exterior derivative. This yields the equation
\beq
d\Sigma^{+}=2idW\wedge\Sigma^{3}_{0}+2id(\phi^{3}\wedge dW),
\eeq
whose complete solution is given by
\beqy
\Sigma^{+}&=&-2idW\wedge\phi^{3}+\Sigma^{+}_{0}+d\phi^{+}
\label{eq:S7}\\
\Sigma^{3}_{0}&=&dW\wedge\phi^{\prime 3},\label{eq:S8}
\eeqy
where $\Sigma^{+}_{0}\equiv Pdy\wedge dz+Qdz\wedge dx+Rdx\wedge
dy$, $\phi^{+}$ is an arbitrary one-form on $\M3$ and
$\phi^{\prime 3}$ is a linear combination of $\{dx,dy,dz\}$.

The final result obtained by synthesizing
eqs.(\ref{eq:S4})(\ref{eq:S5})(\ref{eq:S6})
(\ref{eq:S7}) and (\ref{eq:S8}) is
\beq
\Sigma^{i}J_{i}=\Sigma^{-}_{0}J_{-}+\Sigma^{+}_{0}J_{+}+D\phi,
\eeq
where $\phi\equiv(\phi^{-}-i\phi^{\prime 3})J_{-}+\phi^{3}J_{3}
+\phi^{+}J_{+}$ is an $SL(2,{\bf C})$-valued one-form on $\M3$.
This expression accompanied by the constraint (\ref{eq:S5})
is indeed gauge equivalent to eq.(\ref{eq:TNF}). Thus we
have obtained the desired result.


\appendix{Metric (3.24) violates the strong causality condition.}

Here we will see explicitly that the spacetime metric
(\ref{eq:nullmetric}):
$$
ds^{2}=-dt^{2}-e^{t}dtdZ+e^{2t}d\Xi d\overline{\Xi}
$$
violates the strong causality condition
even if it does not have any closed causal curves.

First we show that this metric at worst has only closed null
curves, namely, it does not contain any closed timelike curves:
\beq
(t(\lambda),Z(\lambda),\Xi(\lambda)):(t_{0},Z_{0},\Xi_{0})
\rightarrow(t_{0},Z_{0}+P_{Z},\Xi_{0}+P_{\Xi}),
\eeq
which satisfy
\beq
(\frac{ds(\lambda)}{d\lambda})^{2}\leq 0\quad\mbox{for}
\quad^{\forall}\lambda\in I\label{eq:inequal}
\eeq
with the inequality holding at least at a point $\lambda_{1}\in I$.
In the above expressions $(P_{Z},P_{\Xi})$ denotes a period
of coordinates $(Z,\Xi)$ on $T^{3}$, and $I$ is some
closed interval in ${\bf R}$. Because $\frac{d\Xi}{d\lambda}$
gives only non-negative contributions to
$(\frac{ds}{d\lambda})^{2}$, $Z$ plays an essential role
for the formation of closed causal curves. So we can replace
$\lambda$ by $Z$. Then a necessary condition for the
existence of closed timelike curves is given by
\beq
-\frac{dt}{dZ}(\frac{dt}{dZ}+e^{t(Z)})\leq 0\quad\mbox{for}\quad
^{\forall}Z\in I_{Z}\equiv[Z_{0},Z_{0}+P_{Z}].\label{eq:ineq2}
\eeq
Let us assume that the inequality holds at $Z_{1}\in I_{Z}$.
Then necessarily $\frac{dt}{dZ}(Z_{1})\neq 0$. Because
$t(Z)$ is by no means a monotonic function, there exist
points $Z_{2}\in I_{Z}$ such that $\frac{dt}{dZ}(Z_{2})<0$
holds. From the inequality (\ref{eq:ineq2}), it follows that,
at all these points, the inequality
$$
e^{t(Z_{2})}+\frac{dt}{dZ}(Z_{2})\leq 0
$$
must hold. This leads to a contradiction because
$\frac{dt}{dZ}(Z_{2})<0$ can be made as close to zero as one likes
while $e^{t(Z_{2})}>0$ is bounded away from zero.

Thus the metric (\ref{eq:nullmetric}) cannot have closed timelike
curves. There can be, however, the case in which the
equality holds in (\ref{eq:inequal}) all along the closed curve,
namely there exist closed null curves.
In this case we must have
$$
\frac{dt}{d\lambda}=\frac{d\Xi}{d\lambda}=0
\quad\mbox{for}\quad^{\forall}\lambda\in I.
$$
For this to be the case, it is necessary and sufficient that the
period with $P_{\Xi}=0$ exists, which is equivalent to the condition
that\footnote{We have set $\xi=\xi_{1}+i\xi_{2}$, $\eta=\eta_{1}+i
\eta_{2}$, and $\zeta=\zeta_{1}+i\zeta_{2}$.}
\beq
(\eta_{1}\zeta_{2}-\zeta_{1}\eta_{2},
\zeta_{1}\xi_{2}-\xi_{1}\zeta_{2},
\xi_{1}\eta_{2}-\eta_{1}\xi_{2})\propto(L,M,N)
\eeq
for some integer-valued vector $(L,N,M)\in{\bf Z}^{3}$.
In other words, the ratios between the components of the vector
in the l.h.s. must be rational.
This does not necessarily hold.
However, because the set of rational numbers are dense in the set of
real numbers, the null line
$(t,Z,\Xi)=(t_{0},Z_{0}+\lambda,\Xi_{0})$ with $0\leq\lambda\leq
\lambda_{M}$ passes through any neighborhood of
$(t_{0},Z_{0},\Xi_{0})$ if $\lambda_{M}$ is taken to be
sufficiently large. The spacetime (\ref{eq:nullmetric}) thus
infringes the strong causality condition, which requires
all the points $p$ in the spacetime to have a neighborhood
which no causal curve intersects more than once.


\end{document}